\documentclass[iop]{emulateapj}

\usepackage{hhline}
\usepackage{natbib}
\usepackage{graphicx}
\usepackage{subfigure}
\usepackage{amsmath}
\usepackage[utf8]{inputenc}
\usepackage{float}
\usepackage [english]{babel}
\usepackage [autostyle, english = american]{csquotes}
\MakeOuterQuote{"}
\usepackage{multirow}
\usepackage{array}
\newcolumntype{M}[1]{>{\centering\arraybackslash}m{#1}}
\newcolumntype{N}{@{}m{0pt}@{}}

\shorttitle{MMR, Exoplanet Detection \& Characterization}
\shortauthors{Tabeshian \& Wiegert}

\def\las{\mathrel{\hbox{\rlap{\hbox{\lower3pt\hbox{$\sim$}}}\hbox{\raise2pt\hbox{$<$}}}}}
\def\gas{\mathrel{\hbox{\rlap{\hbox{\lower3pt\hbox{$\sim$}}}\hbox{\raise2pt\hbox{$>$}}}}}

\begin{document}

\title{Applying a Particle-only Model to the HL Tau Disk}

\author{Maryam Tabeshian\altaffilmark{1} and Paul A. Wiegert\altaffilmark{1,2}}
\affil{$^1$Department of Physics and Astronomy, The University of Western Ontario, London, ON, Canada, N6A 3K7 \\ 
$^2$Centre for Planetary Science and Exploration, The University of Western Ontario, London, ON, Canada, N6A 3K7 \\ \\
\textit{Published in The Astrophysical Journal (ApJ, 857, 3), April 10, 2018}}

\email{mtabeshi@uwo.ca}

\begin{abstract}
Observations have revealed rich structures in protoplanetary disks, offering clues about their embedded planets. Due to the complexities introduced by the abundance of gas in these disks, modeling their structure in detail is computationally intensive, requiring complex hydrodynamic codes and substantial computing power. It would be advantageous if computationally simpler models could provide some preliminary information on these disks. Here we apply a particle-only model (that we developed for gas-poor debris disks) to the gas-rich disk, HL Tauri, to address the question of whether such simple models can inform the study of these systems. Assuming three potentially embedded planets, we match HL Tau’s radial profile fairly well and derive best-fit planetary masses and orbital radii (0.40, 0.02, 0.21 Jupiter masses for the planets orbiting a $0.55~M_\odot$ star at 11.22, 29.67, 64.23 AU). Our derived parameters are comparable to those estimated by others, except for the mass of the second planet. Our simulations also reproduce some narrower gaps seen in the ALMA image away from the orbits of the planets. The nature of these gaps is debated but, based on our simulations, we argue they could result from planet-disk interactions via mean-motion resonances, and need not contain planets. Our results suggest that a simple particle-only model can be used as a first step to understanding dynamical structures in gas disks, particularly those formed by planets, and determine some parameters of their hidden planets, serving as useful initial inputs to hydrodynamic models which are needed to investigate disk and planet properties more thoroughly.\\
\end{abstract}



\section{Introduction}
\label{Sec:Intro4}

Planets are believed to form in protoplanetary disks. While doing so, they create complex symmetric and asymmetric morphological structures. These include density enhancements due to particle trapping in a planet’s pressure bump and mean-motion resonances (MMRs), as well as gap clearing due to dynamical ejection of disk particles as they come into close encounter with the forming planets. In fact, numerical simulations have shown that a planet with only 0.1 Jupiter mass ($M_J$) is capable of pushing the dust away and significantly changing the dust-to-gas ratio of a protoplanetary disk in its vicinity, while a planet mass of at least $1~M_J$ is needed to also form a gap in gas surface density \citep[see for instance,][]{Paardekooper04, Price17}. Such structures provide a wealth of information about the planets that are otherwise difficult to directly observe (such as a planet’s mass), and many studies have attempted to put constraints on the planetary parameters based on how planets affect the distribution of gas and dust in protoplanetary disks \citep[see for instance,][]{Fouchet10, vanderMarel13, Kanagawa15, Kanagawa16}.

Protoplanetary disks are gas-rich (typical gas-to-dust ratio of 100:1 \citep{Collins09} though this changes as they evolve) and a full exploration of their dynamics by numerical methods is expensive in terms of computing power. Such disks also show many features which are similar to those observed in debris disks, i.e., disks of solid particles whose interactions are much easier to model computationally than gas-rich disks. Here we ask the question: how well can the parameters of planets embedded in a protoplanetary disk be extracted using simpler "particle-only" methods? Indeed we find that the masses and radial distances of the planets that may be sculpting the gaps in the HL Tau disk can be extracted with accuracy comparable to that of full hydrodynamic simulations, assuming that there are three hidden planets in the disk. Thus quick, particle-only simulations of protoplanetary disks may be a useful tool for preliminary analyses, and provide useful initial starting points for parameter searches with more complete models.

It should be noted that planet formation is not the only mechanism that is thought to explain the origin of the gap structures in protoplanetary disks. For instance, in a study by \cite{Zhang15},  volatile condensation and rapid pebble growth beyond the snow line are used to reproduce structures such as those observed in the HL Tau disk. On the other hand, secular gravitational instability is also discussed in the literature as one mechanism that could create ring structures in protoplanetary disks \citep[see for instance][]{Takahashi14}. Although these mechanisms may alternatively be used to explain the structures observed in the HL Tau disk, gap opening by planets embedded in this disk remains a strong possibility, and this is what we will consider in the present study. The fact that the eccentricities of HL Tau’s rings increase with increasing distance, and that many of the rings are nearly in a chain of MMRs, indicates that the architecture of the HL Tau disk likely arises from embedded planets \citep[see][]{Brogan15}.

We begin with a description of the literature on the topics of HL Tau’s embedded planets in Section \ref{Sec:HLTau_Studies}, before turning to our own modeling efforts in Section \ref{Sec:Method4} where we discuss our simulations to match the observed intensity profile of the HL Tau disk including the fitting procedure as well as uncertainty measurements. We discuss our results in Section \ref{Sec:Results4}, which includes a comparison to other studies as well as a discussion of MMR gaps in the HL Tau disk. Finally, a summary and conclusions are provided in Section \ref{Sec:SummConc4}.


\subsection{HL Tau Studies to Date}
\label{Sec:HLTau_Studies}

Recent high-resolution observations of a proplanetary disk around the young ($\sim$1 Myr) T Tauri star HL Tauri by the Atacama Large Millimeter/submillimeter Array (ALMA) have revealed unprecedented detailed structures, which are considered likely to be the signatures of planets in the making. This image was taken as part of ALMA’s science verification phase in 2014 October and was released a month later \citep[see][]{NRAO14}. The disk was observed in dust continuum emission at 233 GHz (1.28 mm) using 25–30 antennas and a maximum baseline of 15.24 km as part of ALMA’s Long Baseline Campaign, and achieved an angular resolution of 35 mas, equivalent to 5 AU at HL Tau’s distance of 130 pc. It reveals a series of concentric gaps that have become the subject of many studies, shedding light on the properties of the planets that are believed to be carving out these gaps, and providing a better understanding of the processes involved in the formation and evolution of planets and planetary systems. Table \ref{Tab:HLTau_params} lists orbital radii and the masses of the potentially hidden planets in five of the seven gaps that can be identified in the ALMA image, derived from past studies that used hydrodynamic and numerical simulations as well as analytic estimates which we briefly review here. We will compare the planetary masses derived in the literature to the results of our "particle-only" model in Section \ref{Sec:Results4}.

\cite{Brogan15} identified seven pairs of distinct dark and bright rings in the ALMA image of the HL Tau disk which they labeled D1...D7 and B1...B7 (more on this in Section \ref{Sec:SummConc4}). They approximated the radial distance of the center of each ring by making a cross-cut along the disk’s major axis and found the dark rings to be at $13.2\pm0.2$, $32.3\pm0.1$, $\sim42$, $\sim50$, $64.2\pm0.1$, $73.7\pm0.1$, and $\sim91.0$ AU, placing the first four dark rings in a chain of MMRs, specifically 1:4:6:8. \cite{Pinte16} measured the missing dust mass in each of the seven gaps by integrating the dust surface density of each gap and comparing it to its surrounding bright rings. They argued that these provide the mass of the rocky core of the possibly embedded planet in each dark ring.

Other authors have also attempted to constrain the masses of the planets that are believed to be shepherding the HL Tau gaps. Based on the depth of the gap seen $\sim$ 30 AU from a central star, \cite{Kanagawa15} estimated the mass of its embedded planet. They did so by using the relationship between the depth of a gap formed by a planet in its feeding zone in a protoplanetary disk and the mass of the planet as well as the disk’s viscosity and scale height \citep[see]{Duffell13, Fung14, Kanagawa15}, given by \citep{Kanagawa15}:

\begin{equation}
\label{Eq:Depth_Mass}
\frac{M_p}{M_\star} = 5\times10^{-4} \left(\frac{1}{{\Sigma_p}/{\Sigma_0}}-1\right)^{1/2} \left(\frac{h_p}{0.1}\right)^{5/2} \left(\frac{\alpha_{ss}}{10^{-3}}\right)^{1/2} ~,
\end{equation}

\noindent where $M_p$ is the planet's mass in stellar mass units $M_\star$, ${\Sigma_p}/{\Sigma_0}$ is the gap depth which is the ratio of the surface density of the planet-induced gap to that of the unperturbed disk, $h_p$ is the disk's aspect ratio at the planet's orbital radius ($h/r$, with $h$ being the scale height), and $\alpha_{ss}$ is the Shakura-Sunyaev kinematic viscosity parameter \citep{Shakura73}. 

Adopting a stellar mass of $1.0 ~M_\odot$, a viscosity parameter of $10^{-3}$, and estimating the gap depth and the disk's aspect ratio to be $\sim 1/3$ and $\sim 0.07$ respectively, \cite{Kanagawa15} were able to determine that the mass of the planet at 30 AU is at least $0.3 ~M_J$. Also using the gap depth (Equation \ref{Eq:Depth_Mass}) and a method based on angular momentum transfer analysis in gas disks, \cite{Akiyama16} estimated the masses of the planets in the HL Tau system to be comparable to or less than 1 $M_J$.

Estimating the dust mass deficits in the gaps as done by \cite{Pinte16}, \cite{Kanagawa15}, and \cite{Akiyama16} provides a lower limit for the planet masses since an accurate measurement of the gap depth requires high signal-to-noise ratio data, otherwise the gaps cannot be fully resolved and seem to be partially "filled in" \citep{Pinte16}. For this reason, in a follow-up paper, instead of using the gap depth to measure the masses of the planets, \cite{Kanagawa16} derived an empirical relationship between the width of a planet-induced gap and planet mass, disk aspect ratio, and viscosity. Using two-dimensional hydrodynamic simulations, and assuming that the dust particles are strongly tied to the gas (i.e., dust filtration is not a major concern), they determined this relationship to be:

\begin{equation}
\label{Eq:Width_Mass}
\frac{M_p}{M_\star} = 2.1\times10^{-3} \left(\frac{\Delta_{gap}}{r_p}\right)^{2} \left(\frac{h_p}{0.05 ~ r_p}\right)^{3/2} \left(\frac{\alpha_{ss}}{10^{-3}}\right)^{1/2} ~,
\end{equation}

\noindent with $r_p$ and $\Delta_{gap}$ being the orbital radius of the planet and the width of the gap it creates in the disk, respectively, where the gap edges are defined by regions where the surface density drops to less than half the unperturbed surface density.

Using Equation \ref{Eq:Width_Mass} to determine planetary masses probably results in more accurate estimates than simply using the size of each planet’s Hill radius, which tends to predict rather large planetary masses. \footnote{The Hill radius, $r_H$, is defined as $(\frac{M}{3 M_*})^{(1/3)} ~a$, with $M$ and $M_*$ the masses of the planet and the star, respectively, and $a$ the semimajor axis of the planet's orbit \citep{Murray99}. $r_H$ defines the region around a planet where its gravity dominates over that of the star: systems of moons, for example, must reside well within a planet’s Hill sphere to be stable.} For the planets in the HL Tau disk, \cite{Akiyama16} measured the gap widths to be 5.0, 4.1, 6.2, and 4.5 AU, at $r_p$ = $\sim$13.5, $\sim$32.4, $\sim$65.2, and $\sim$77.2 AU and used the size of each planet's Hill sphere to calculate its mass. This resulted in planetary masses of $88.8^{+5}_{-5}$, $3.6^{+0.7}_{-0.6}$, $1.5^{+0.5}_{-0.5}$, and $0.3^{+0.1}_{-0.1}$ $M_J$, much larger than other mass estimates for the HL Tau planets, especially the innermost planet whose Hill radius suggests a stellar-mass body. Although the possibility of low-mass stellar companions in this system is not ruled out, high-sensitivity direct imaging in the mid-infrared by \cite{Testi15} did not reveal any point sources in the HL Tau disk. Although their observations were more focused around the gaps at $\sim$ 70 AU (for which the contrast level reached was $\sim$ 7.5 mag.), their search for point sources in the HL Tau disk was not exclusive to the outer disk. Nevertheless, to examine the possibility of stellar/substellar companions in the HL Tau disk, further studies are needed to determine the stability of the system under such conditions. Therefore, we exclude mass measurements from planetary Hill radii when we later compare our results to those of others (see Section \ref{Sec:Compare}).

It is worth noting here that dust filtration by the planet’s pressure bump as well as dust migration under radiation and drag forces can also cause gaps to be filled in temporarily and yield inaccurate measurements of the gap width and depth. Thus the mass of a gap-opening planet derived from the width and depth of the gap must be taken with caution, particularly in cases where the disk is massive and hence there is high rate of collisional fragmentation down to grain sizes that are affected by radiation and drag forces. However, such effects are more important when planetary masses are estimated from the depths of the gaps than their widths \citep[see][]{Dong15}.

Using hydrodynamic simulations and radiative transfer models, \cite{Jin16} attempted to match the width and depth of the three prominent gaps in the HL Tau disk, located at at 13.1, 33.0, and 68.6 AU and constrained the masses of the planets that are believed to be in those gaps to be 0.35, 0.17, and 0.26 $M_J$, respectively, assuming no planet migration through the disk. The model assumes a disk mass of $\sim 7.35\times10^{-2} ~M_\odot$ and the same $\alpha_{ss}$ parameter as \cite{Kanagawa15} while the dust-to-gas ratio is taken to be $1\%$. Furthermore, the authors also tried to match the eccentricities of the gaps where they placed the three planets and found them to be 0.246, 0.274, and 0.277, respectively. On the other hand, smoothed particle hydrodynamic (SPH) models by \cite{Dipierro15} constrained the masses of the planets embedded in the HL Tau disk to be 0.2, 0.27 and 0.55 $M_J$ with planets at 13.2, 32.3 and 68.8 AU.

Gas and dust interact differently with planets. Numerical simulations by \cite{Jin16} showed that the three gaps formed by tidal interaction with the embedded planets in the HL Tau disk are shallower in gas distribution and deeper in dust, though both have similar morphologies (see their Figure 1). The difference in the gap’s gas and dust surface density arises from the fact that submillimeter dust is pushed toward the edges of the gap as it starts to open since gas drag tends to accumulate dust particles in high-pressure regions as suggested by the enhanced dust emission near gap edges \citep[see][]{Haghighipour03, Fouchet07, Maddison07}. 

Most authors place three planets in the HL Tau disk; however, the possibility of additional planets in this disk has also been discussed in the literature. For instance \cite{Tamayo15} considered the possibility of up to five planets in the HL Tau disk at nominal radii of 13.6, 33.3, 65.1, 77.3, and 93.0 AU. This places the outer three planets nearly in a chain of 4:3 MMR. The authors determined the masses of the five planets under two different scenarios: if the planets are not in MMR, they found a maximum mass of $\sim 2$ Neptune masses for the outer three bodies; however, if the outer three planets are in resonance, as suggested by the locations of the gaps, they can grow to larger masses via resonant capture as they migrate through the disk during which their masses can reach at least that of Saturn. The masses of the two inner planets were not well constrained in this study since these planets are dynamically decoupled from the other three. 

Planets forming in a multi-planet system can grow to where the system becomes unstable simply because of the growth in the sizes of the planets' Hill spheres. Planets whose orbits around the star are separated by less than several of their mutual Hill spheres are unstable: this stability criterion is defined by \cite{Gladman93} who suggested planets that are separated by less than $3.46 ~r_H$ destabilize on a timescale that is roughly their conjunction period. On the other hand, \cite{Tamayo15} showed through numerical simulations, using the REBOUND package \citep{Rein12}, that planets can still survive well beyond the above stability criterion if they capture in MMR at low masses and grow together. This is because resonance mitigates the effect of close encounters. For the system to be Hill stable, the maximum masses for the three outermost planets in the HL Tau disk were found by \cite{Tamayo15} using Equation \ref{Eq:M_crit}, taking the stellar mass to be $M_\star=0.55 ~M_\odot$:

\begin{equation}
\label{Eq:M_crit}
M \lesssim M_{crit} = 8~\left(\frac{M_\star}{M_\odot}\right) \left(\frac{\Delta a/a}{0.1}\right)^3 ~M_\oplus ~,
\end{equation}

\noindent where $\Delta {a}$ is the planet separation and $M_{crit}$ is the maximum mass to ensure stability.

Therefore, according to numerical simulations of the HL Tau disk by \cite{Tamayo15}, if the outer three planets are not in MMR they become unstable at conjunction timescale once they exceed the mass threshold beyond which their separation becomes less than $\sim 3.5 ~r_H$. However, if they are captured at resonances while they migrate through the disk, they can grow well past the above limit until they become so massive ($\sim$ 40\% beyond mass of Saturn or $0.44 ~M_J$) that their mutual gravitational perturbation at conjunctions brings them out of resonance at which point swift instability ensues \citep{Tamayo15}. 

\begin{table*}[ht]
\scriptsize
\centering
\caption{Estimated Orbital Radii and Masses of Possible Planets Forming HL Tau’s Five Major Gaps from the Literature}
\label{Tab:HLTau_params}
\begin{tabular}{|m{0.95cm}|m{0.95cm}|m{0.95cm}|m{0.95cm}|m{0.95cm}|m{0.95cm}|m{0.95cm}|m{0.95cm}|m{0.95cm}|m{0.95cm}|m{0.5cm}|m{1.7cm}|m{1.1cm}|}
\hline
\multicolumn{5}{|c|}{\multirow{3}{*}{\normalsize{\textbf{r$_p$ ($AU$)}}}}  & \multicolumn{5}{c|}{\multirow{3}{*}{\normalsize{\textbf{M $(M_J)$}}}}  & \multicolumn{1}{c|}{\multirow{3}{*}{\normalsize{\textbf{\!\!\!M$_\star (M_\odot)$\!\!}}}} & \multicolumn{1}{c|}{\multirow{3}{*}{\normalsize{\textbf{Method}}}}     & \multicolumn{1}{c|}{\multirow{3}{*}{\normalsize{\textbf{Ref.}}}}  \\ [15pt] \cline{1-13}

\begin{tabular}[c]{@{}l@{}} \\ \\ $11.2^{+0.2}_{-0.1}$ \\ \\ \\ \end{tabular} & $29.7^{+2.9}_{-2.9}$ & $64.2^{+0.3}_{-0.3}$ & & & 
$0.40^{+0.02}_{-0.00}$ & $0.02^{+0.03}_{-0.02}$ & $0.21^{+0.02}_{-0.01}$ & & & 
~~~~0.55 &
\begin{tabular}[c]{@{}l@{}}Particle-only \\numerical\\ sim. (Wisdom-\\Holman)\end{tabular} & This work \\ \hline

\begin{tabular}[c]{@{}l@{}} \\ \\ 13.1 \\ \\ \\ \end{tabular} & 33.0 & 68.6 & & & 
0.35 & 0.17 & 0.26 & & & 
~~~~0.55 &
HD \& radiative transfer & \cite{Jin16} \\ \hline

\begin{tabular}[c]{@{}l@{}} \\ \\ 11.8 \\ \\ \\ \end{tabular} & 32.3 & 82 & & & 
0.77 & 0.11 & 0.28 & & & 
~~~~0.55 &
2-D\! HD sim., Gap width measurements (Eq.\!\ref{Eq:Width_Mass}) & \multirow{2}{*}{\begin{tabular}[c]{@{}l@{}} \!Kanagawa \\ \!et al. \\ \!(2016) \end{tabular}} \\[15pt] \hline

\begin{tabular}[c]{@{}l@{}} \\ \\ $13.5^{+0.4}_{-0.4}$ \\ \\ \\ \end{tabular} & $32.4^{+0.6}_{-0.4}$ & $65.2^{+1.3}_{-0.9}$ & $77.2^{+0.8}_{-0.7}$ & & 
$>\!0.85$ & $>\!0.61$ & $>\!0.62$ & $>\!0.51$ &   & 
~~~~0.55 & 
Gap depth measurements (Eq.\!\ref{Eq:Depth_Mass})  & 
\multirow{2}{*}{\begin{tabular}[c]{@{}l@{}} \!Akiyama \\ \!et al. \\ \!(2016) \end{tabular}} \\[15pt] \hline

$13.6^{+0.2}_{-0.2}$ & $33.3^{+0.2}_{-0.2}$ & $71.2^{+0.5}_{-0.5}$ & $93.0^{+0.9}_{-0.9}$ & & 
& & \begin{tabular}[c]{@{}l@{}}$\lesssim 0.30$\\\!\!No \!MMR\!\!\\           \\ $\gtrsim 0.72$\\MMR\end{tabular} & 
\begin{tabular}[c]{@{}l@{}}$\lesssim 0.30$\\\!\!No \!MMR\!\!\\           \\ $\gtrsim 0.72$\\MMR\end{tabular} & & 
\multirow{2}{*}{\begin{tabular}[c]{@{}l@{}} \\ \\~~~0.55 \\ \end{tabular}} & \multirow{2}{*}{\begin{tabular}[c]{@{}l@{}}N body sim., \\ REBOUND \\package \\ (Rein \& Liu, \\ 2012) \end{tabular}} & 
\multirow{2}{*}{\begin{tabular}[c]{@{}l@{}} Tamayo \\ et al. \\ (2015) \end{tabular}} \\ \cline{1-10}

$13.6^{+0.2}_{-0.2}$ & $33.3^{+0.2}_{-0.2}$ & $65.1^{+0.6}_{-0.6}$ & $77.3^{+0.4}_{-0.4}$ & $93.0^{+0.9}_{-0.9}$ & 
& & \begin{tabular}[c]{@{}l@{}}$\lesssim 0.11$\\\!\!No \!MMR\!\!\\           \\ $\lesssim 0.30$\\MMR\end{tabular} & 
\begin{tabular}[c]{@{}l@{}}$\lesssim 0.11$\\\!\!No \!MMR\!\!\\           \\ $\lesssim 0.30$\\MMR\end{tabular} & 
\begin{tabular}[c]{@{}l@{}}$\lesssim 0.11$\\\!\!No \!MMR\!\!\\           \\ $\lesssim 0.30$\\MMR\end{tabular} & 
& & \\ 
\hhline{|=|=|=|=|=|=|=|=|=|=|=|=|=|}

\begin{tabular}[c]{@{}l@{}} \\ \\ $11.3^{+0.2}_{-0.1}$ \\ \\ \\ \end{tabular} & $29.4^{+2.6}_{-3.6}$ & $63.7^{+0.5}_{-0.4}$ & & & 
\!\!\!$0.81^{+0.02}_{-0.01}$ & \!\!\!$0.04^{+0.04}_{-0.04}$ & \!\!\!$0.37^{+0.01}_{-0.03}$ & & & 
~~~~1.3 &
\begin{tabular}[c]{@{}l@{}}Particle-only \\numerical\\ sim. (Wisdom-\\Holman)\end{tabular} & This work \\ \hline

\begin{tabular}[c]{@{}l@{}} \\ \\ 13.2 \\ \\ \\ \end{tabular} & 32.3 & 68.8 & & & 
0.2  & 0.27 & 0.55 & & & 
~~~~1.3 &
3-D dust+gas SPH & \multirow{2}{*}{\begin{tabular}[c]{@{}l@{}} \!Dipierro \\ \!et al. \\ \!(2015) \end{tabular}} \\[15pt] \hhline{|=|=|=|=|=|=|=|=|=|=|=|=|=|}

\begin{tabular}[c]{@{}l@{}} \\ \\ 11.8 \\ \\ \\ \end{tabular} & 32.3 & 82 & & & 
1.4 & 0.2 & 0.5 & & & 
~~~~1.0 &
2-D\! HD sim., Gap width measurements  (Eq.\!\ref{Eq:Width_Mass}) & \multirow{2}{*}{\begin{tabular}[c]{@{}l@{}} \!Kanagawa \\ \!et al. \\ \!(2016) \end{tabular}} \\[15pt] \hline

\begin{tabular}[c]{@{}l@{}} \\ \\ $13.2^{+0.2}_{-0.2}$ \\ \\ \\ \end{tabular} & $32.3^{+0.1}_{-0.1}$ & $64.2^{+0.1}_{-0.1}$ & $73.7^{+0.1}_{-0.1}$ & $91$ & 
\begin{tabular}[c]{@{}l@{}}$\!\!\!\!>\!\!0.02^{+0.01}_{-0.01}$\end{tabular} & \begin{tabular}[c]{@{}l@{}}$\!\!\!\!>\!\!0.07^{+0.01}_{-0.01}$\end{tabular} & \begin{tabular}[c]{@{}l@{}}$\!\!\!\!>\!\!0.03^{+0.00}_{-0.01}$\end{tabular} & \begin{tabular}[c]{@{}l@{}}$\!\!\!\!>\!\!0.08^{+0.03}_{-0.05}$\end{tabular} & \begin{tabular}[c]{@{}l@{}}$\!\!\!\!>\!\!0.11^{+0.03}_{-0.06}$\end{tabular} & ~~~~1.7 & \begin{tabular}[c]{@{}l@{}}\!\!Surface density\\measurements\end{tabular} & \multirow{2}{*}{\begin{tabular}[c]{@{}l@{}} \!Pinte \\ \!et al. \\ \!(2016) \end{tabular}} \\[15pt] \hline

\end{tabular}
\\[10pt]
\begin{flushleft}
\textbf{Note:} If the two gaps at D5 and D6 are formed by a single planet at 71.2 AU, \cite{Tamayo15} estimated the masses of the two outermost planets to be $\lesssim 0.30 ~M_J$ if the two planets at D5+D6 and D7 are not in MMR and $\gtrsim 0.72 ~M_J$ if they are. In the four-planet scenario, \cite{Pinte16} determined the mass of a single planet at 69.0 AU forming the gap at D5+D6 to be at least $0.44^{+0.05}_{-0.09} ~M_J$, though they used a much larger stellar mass to derive those planetary masses. \cite{Kanagawa16} used two different stellar masses in their measurements of planet masses based on gap widths, both of which are listed here. Their mass measurement for the planet orbiting around a 1 $M_\odot$ star at $\sim$ 30 AU is consistent with their earlier paper \citep{Kanagawa15} where they had determined the mass from gap depth to be at least 0.3 $M_J$ using the same stellar mass. This table also lists the results of our work, determined for two stellar masses of 0.55 and 1.3 $M_\odot$, explained in subsequent sections.
\\[40pt]
\end{flushleft}
\end{table*}

Moreover, their numerical simulation suggested that the system would be substantially more stable if not all the gaps were made by planets, particularly the more closely spaced gaps at 64.2 and 73.7 AU, or D5 and D6 according to \cite{Brogan15} (Note that these two gaps are at 65.1 and 77.3 AU in \cite{Tamayo15}). They suggested that these two gaps might not be made by two different planets; there might instead be a single planet at 71.2 AU that has shaped a horseshoe-like gap in the disk of HL Tau. If four planets are considered instead of five in the HL Tau system, their numerical simulations put a final mass limit of at least $230 ~M_\oplus$ for the outer two planets if they are in MMR while they can reach $\sim$1 Saturn mass if they are not.

Other authors have also suggested that the double gap at D5 and D6 in the HL Tau disk could be made by a single planet exciting Lindblad torques with the bump in the middle being co-orbital horseshoe material where the planet is possibly hiding. Using 2D gas+dust hydrodynamic simulations combined with radiative transfer modeling, \cite{Dong17} showed how super-Earths placed in a low-viscosity disk can produce characteristic double gaps in mm-dust distribution and argued that the D5+D6 gaps in the HL Tau disk could be carved by a single planet. Besides the double gap on either side of the planet’s orbit, their simulations also suggested that additional gaps could arise in the disk for a single planet, depending on disk and planet parameters. In a more recent paper by \cite{Bae17}, using 2D hydrodynamic simulations of the gas component of the HL Tau disk, the authors were able to reproduce not only D5 and D6 with a single planet at 68.8 AU, but also noticed that the same planet can reproduce the D1 and D2 gaps at 13.2 and 32.3 AU, though their model did not reproduce the finer structures such as D3, D4, and D7. They also argued that the mass of the planet can be constrained from the positions of multiple gaps, provided that the disk temperature profile can be accurately measured.

Table \ref{Tab:HLTau_params} summarizes the masses and locations of the possible planets in the HL Tau system obtained by the studies mentioned above. It is important to note that the masses derived for the HL Tau planets in the literature depend on the mass of the central star, which is not well constrained. Estimates of HL Tau’s stellar mass range from 0.55 $M_\odot$ \citep[e.g.,][]{Tamayo15} to 1.7 $M_\odot$ \citep[e.g.,][]{Pinte16}. Therefore, in order to be able to compare our results to those of others, we only focus on two previously used values of 1.3 and 0.55 $M\odot$. Nevertheless, even for the same stellar mass, Table \ref{Tab:HLTau_params} shows that, despite many attempts to constrain planetary parameters in the HL Tau disk, much work is still needed to determine the number and parameters of its potentially embedded planets. This is the primary motivation of this work: can we reproduce the key features of the HL Tau disk using the computationally inexpensive model of a particle-only disk to address whether some parameters of its planets, specifically their mass and orbital radii, can be determined without the need for sophisticated models which are, nevertheless, required to fully describe gas-rich disks? In Table \ref{Tab:HLTau_params}, we also list our results for comparison with others but will explain how we arrived at these values in the next two sections.


\section{Method}
\label{Sec:Method4}

\subsection{The HL Tau Disk Profile} 
\label{Sec:profile4}

The observed profile of the HL Tau disk used here is extracted by the following method. First, we obtained the FITS image of the HL Tau disk available publicly at the ALMA website and observed in dust continuum emission in band 7 (the highest resolution). We made a cross-cut across the disk’s major axis to extract HL Tau’s radial brightness profile in Jy/beam per radial distance from the star. The extracted profile is 186 pixels long over a physical distance of 115 AU. However, the resolution of the image is only 35 mas or $\sim$5 AU at 130 parsec \citep[see][]{NRAO14} and so we assess that we really only have $115/5 \approx 23$ bins for the purposes of determining our degrees of freedom (see Section \ref{Sec:uncertainties4}) and $186/23 \approx 8$ pixels per bin.


\subsection{Simulations}
\label{Sec:Sims4}

Our simulations are performed with a symplectic integrator based on the Wisdom–Holman algorithm \citep{Wisdom91}. A fixed timestep of 150 days is used for all simulations. Only point particles are simulated, without any gas drag, radiation pressure, or Poynting–Robertson drag. These effects are likely to be important in sculpting the HL Tau disk but our purpose here is to determine what, if any, of the planetary parameters can be recovered by the simplest possible model.

Simulations are run for 10,000 yr ($\sim 1000$ inner orbits) and recorded at 100 yr intervals. Three planets and 1000 particles are placed within the disk on circular orbits around a 1.3 or 0.55 solar-mass central star. Particles are removed if they reach a distance less than $\sim$500 solar radii or greater than 220 AU. The planets are placed nominally at 11.7, 29.1, and 64.5 AU based on the locations of the gaps in the HL Tau disk, but the planets' locations will be varied as part of the fitting process, described in Section \ref{Sec:Fitting4}.

Simulated disk profiles are created from the last five snapshots of the disk. The use of several snapshots increases our signal-to-noise without the computational expense associated with simulating additional particles, though it assumes that the disk is in a quasi-steady state. Examination of the disk during the final stages confirms that indeed the disk structures are well-established.

For plotting purposes, the simulation data are extracted into a histogram with 186 bins to match the observations. The bins are weighted by the blackbody emission of their particles assuming a dust albedo of 0.5 and emissivity of 1.0 at mm wavelengths to calculate the equilibrium temperature of the disk particles. The stellar luminosity and effective temperature are also uncertain but are taken to be $8.3 ~L_\odot$ and 4000 K, respectively \citep{Ruge16}. For calculation of the $\chi^2$ of our fits, the data are box-car smoothed down to the effective resolution of the observations (8.3 bin box-car). On the basis of the $\chi^2$ value, new parameter values are chosen and a new simulation is initialized. The whole process is iterated until convergence is achieved.


\subsection{Fitting} 
\label{Sec:Fitting4}

Best-fit parameters are established on the basis of the $\chi^2$ between the observational profile and a simulated profile normalized to the first bin in the observed profile. This normalization reduces our degrees of freedom by one. Minimization of the $\chi^2$ parameter is accomplished using Interactive Data Language (IDL) and the Amoeba package, which is a multi-dimensional derivative-free optimization algorithm based on the downhill simplex method of \cite{Nelder64}. Typical Amoeba runs require 900–1000 simulations and a total of 10 hr to complete on a single CPU. Amoeba requires the tolerance to be at least equal to the machine’s double precision, so we set the tolerance to $10^{-12}$. This is the decrease in the fractional value of the $\chi^2$ in the terminating step.

Chi-squared minimization using the Amoeba algorithm does not require calculating derivatives. Furthermore, each iteration only takes one or two function evaluations and therefore Amoeba converges faster than some other minimization routines such as nonlinear least-squares fitting using the Levenberg–Marquardt algorithm \citep{Marquardt44, Levenberg44} which takes several calculations per iteration. Amoeba is also more robust for problems with stochastic components such as what we are dealing with here (e.g., the particle positions are chosen randomly for each simulation, which introduces some statistical noise to the radial profiles), and we chose the Amoeba algorithm for these reasons.

A downside to using Amoeba is that it can get to a point where the changes in the parameter values become insignificant before a minimum is reached. Thus it is generally recommended to restart Amoeba from the point where it claims to have found a minimum \citep[see][]{Press92} and this is what we do 10 times until the routine converges again. Our procedure was to first perform initial minimization runs using parameter values chosen arbitrarily, except for the orbital radii of the three planets (these were estimated from the locations of the major gaps in the HL Tau disk) and each parameter was allowed to vary by $\pm 50\%$ by the minimization routine. From the lowest $\chi^2$ obtained from these initial runs (our "initial solution"), in order to ensure as much as possible that the minimum $\chi^2$ achieved is the global minimum, we performed 10 additional minimization runs where we changed the initial conditions such that each parameter fell randomly within $10\%$ of that obtained from the initial solution. At the end, we recorded the parameters that produced the lowest $\chi^2$ from the $10+1$ Amoeba runs. Our restarting process helps avoid terminating at a spurious local minimum, but we cannot exclude the possibility of a true global minimum that might exist far away from our final result in parameter space.

For our simulations here, we fit 10 parameters of the planets and disk (in our model with the broken power-law but seven when we use a single power-law for disk density distribution, see Section \ref{Sec:Results4}). We assume that there are three planets on circular orbits. In addition to the masses and orbital radii of these three planets, we also fit a power law to the disk surface density. The surface density of circumstellar disks is generally taken to have a profile of the form $\Sigma \propto R^{-\alpha}$ with the power-law index, $\alpha$, between 0 and 1 depending on the mass of the protoplanetary disk \citep{Andrews07b} (Note that the power-law index derived from Minimum Mass Solar Nebula is 1.5 \citep{Weidenschilling77}). However, the use of a single power law does not well reproduce the radial profile of HL Tau’s flux density. A much lower $\chi^2$ value is obtained by selecting a different power-law index beyond the location of the outermost planet (see Section \ref{Sec:Results4}). \cite{Yen16} also used a broken power law in their measurements of gap widths and depths in the HL Tau disk where the slopes of the dust distribution based on the column density of HCO$^+$, assuming that gas and dust are well coupled thermally, were found to be $0.5\pm0.2$ at $\sim 20$ AU and $0.9\pm0.3$ at $\sim 60$ AU, suggesting a steep decline in dust continuum emission beyond where the outer major gap lies. \cite{Jin16} proposed that the deficit in dust in the outer part of the HL Tau disk is due to the inward drift of dust caused by gas drag and the absence of a source to supply the dust at large radii \citep[also see][]{Birnstiel14}. In fact, disks are found to have exponentially tapered edges and an exponential decrease in dust surface density has also been observed for a number of other circumstellar disks \citep[see for instance,][]{McCaughrean96} with power-law indices beyond the above-mentioned range, suggesting that the pure power-law relation (i.e., $\Sigma \propto r^{-\alpha}$) does not accurately represent a disk’s intensity profile and must be replaced by an exponentially truncated density distribution with $\Sigma \propto r^{-\gamma}$, where $\gamma$ is the exponent in the viscosity dependence on distance from the star \citep[e.g.,][]{Hartmann98}.  For simplicity (that is, to avoid adding additional parameters to our fit), we assume a standard power-law slope without an exponential term. Therefore, we argue that our fit to the radial profile of the outer disk would be improved if we adopted the above surface density profile and incorporated dust re-generation and gas drag in our model, which we leave to future work.


\subsection{Uncertainties} 
\label{Sec:uncertainties4}

Uncertainties in the fitted parameters are estimated based on the $\chi^2$ values. The number of degrees of freedom, $\nu$, will be the effective number of bins (23, see Section \ref{Sec:profile4}) minus one for the normalization discussed in Section \ref{Sec:Fitting4}, and minus one for each free parameter. We have 10 free parameters, giving us a total of 12 degrees of freedom.

The uncertainties to be at the locations in phase-space can be approximated as where the $\chi^2$ value is increased over its minimum value by an amount $\Delta \chi^2$ dependent on $\nu$ and the stringency of the uncertainty bounds desired. Here we choose a $p=0.95$ (nominally $2\sigma$) confidence region, which means that our uncertainties correspond to the locations for which \citep{Press92}:

\begin{equation}
\label{Eq:inc_gamma}
Q \left(\frac{\nu}{2}, \frac{\Delta \chi^2}{2} \right) = 1-p ~,
\end{equation}

\noindent where $Q$ is the incomplete gamma function, and $\Delta \chi^2$ gives the increase in $\chi^2$ corresponding to our uncertainty.

Note that we compute our uncertainties from $\chi^2$ values with all the parameter values except the one in question held constant. This implicitly assumes that the parameters are uncorrelated, which we assume here for reasons of simplicity and practicality. Our $\chi^2$ is derived by a process with inherent stochasticity (i.e., the initial conditions of particles within the disk have a random component), thus we have too many free parameters and too noisy a system to determine the covariance between them all effectively. This will be more apparent when the uncertainty results are discussed in Section \ref{Sec:Results4}.


\section{Results} 
\label{Sec:Results4}

\begin{figure*}
    \centering
    \includegraphics[width=18cm,height=5.5cm]{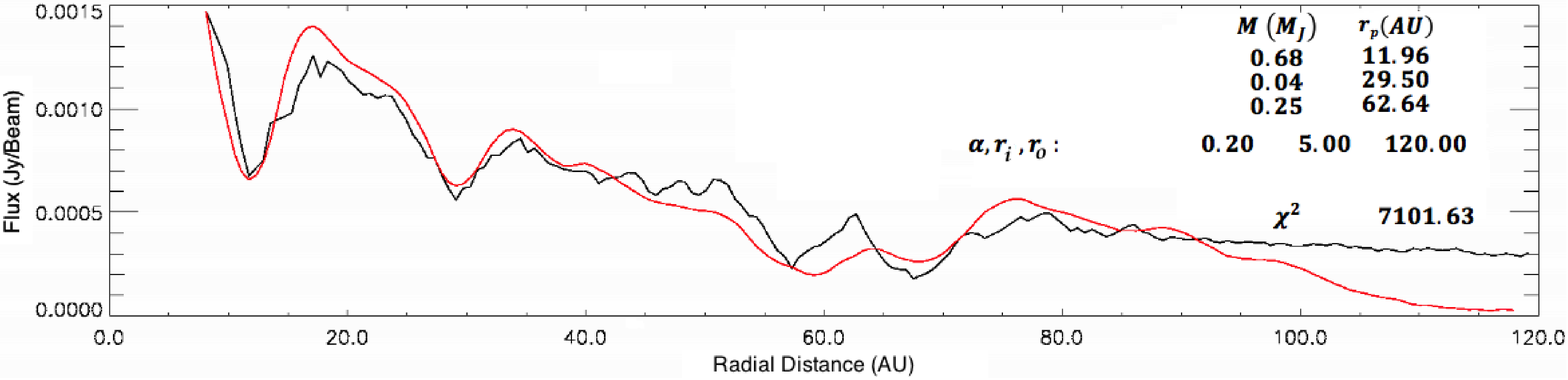}
    \caption[Comparison between the radial profile of the HL Tau disk observed by ALMA and our simulation using a single power-law profile drawn from our final best-fit values.]{Comparison between the radial profile of the HL Tau disk extracted from the FITS image observed by ALMA at band 7 (red) and our simulation drawn from our final best-fit values for a disk with a single power law (black). We place three planets at nominal radii of 11.03, 28.91, and 64.52 AU around a $1.3 ~M_\odot$ star and allow Amoeba to determine the best-fit parameters (i.e., the three planet masses, $M(M_J)$, and orbital radii, $r_p ~(AU)$, as well as the power-law index, $alpha$) by minimizing the $\chi^2$. For simplicity, we assume that the three planets are in circular orbits but acknowledge that the gaps in the HL Tau disk are found to have some eccentricity (see Section \ref{Sec:HLTau_Studies}). Here we use a single power-law surface density index for the disk that extends from $r_i=5.0$ AU to $r_o=120.0$ AU. However, the model with a single power-law index fails to reproduce the disk profile well beyond the location of the outermost planet.}
    \label{Fig:1alpha}
\end{figure*}

As mentioned in Section \ref{Sec:Fitting4}, a single power-law index for the surface density cannot reproduce the observed density profile of the HL Tau disk due to a steep fall-off in the outer part of the disk beyond the location of the outermost planet. This is shown in Figure \ref{Fig:1alpha}. Therefore, we break the disk into two segments, each having a different power-law index, $\alpha_1$ and $\alpha_2$, which we leave as free parameters in our simulations. We also allow the location of the boundary between the two segments to vary, and introduce an additional parameter to allow for a change in the surface density of the disk at the boundary between the two segments.

\begin{figure*}
    \centering
    \includegraphics[width=18cm,height=5.5cm]{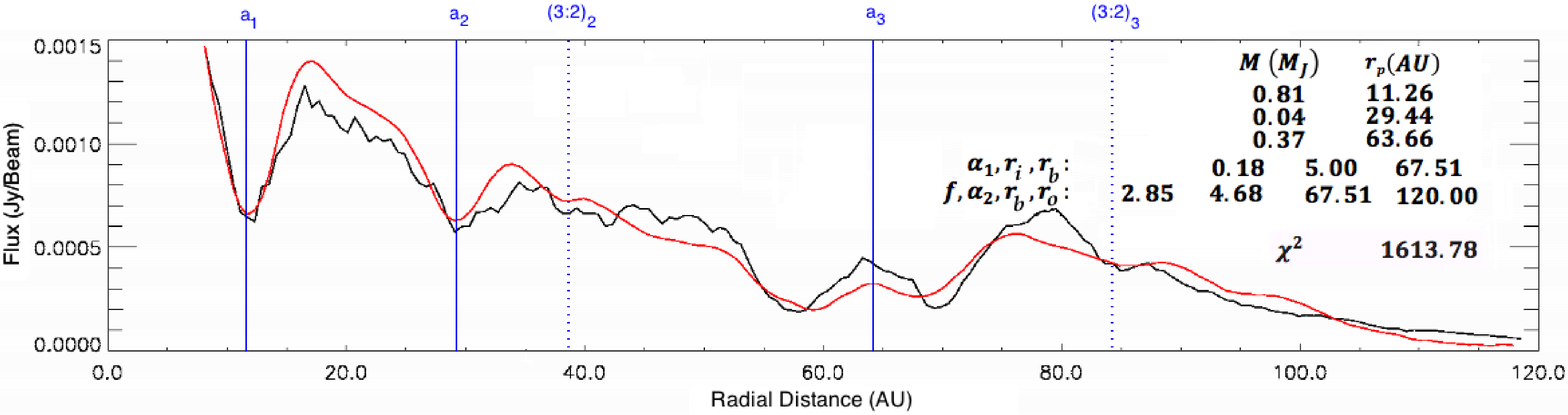}
    \caption[Same as Figure \ref{Fig:1alpha} but with two different power-law indices for dust surface density distribution with a slope break beyond the orbit of the outermost planet.]{Same as Figure \ref{Fig:1alpha} except that we use two different power-law indices for dust surface density distribution to account for the exponentially decaying surface density profile of the HL Tau disk and the steeper slope beyond the orbit of the planet that we place inside the third major gap. The $\chi^2$ value in this case is significantly improved. The nominal locations of the three planets are shown by the solid blue lines. We also identify two gaps that fall at MMRs with the planets at $r_{p_2}$ and $r_{p_3}$ and mark their locations with dotted blue lines (see Section \ref{Sec:HLTau_MMR} for a discussion on possible MMR gaps in the HL Tau disk).}
    \label{Fig:2alpha}
\end{figure*}

To obtain the best-fit values, we thus need to include 10 free parameters in our simulations: one for each planet’s mass ($M$) and orbital radius ($r_p$), two for the differential surface density power-law indices ($\alpha_1$ and $\alpha_2$), one for the transition point that separates the two parts of the disk with different slopes ($r_b$), and finally one for the fractional increase in surface density at the transition point ($f$). Note that we keep $r_i$ and $r_o$ fixed at 5.0 and 120.0 AU which roughly mark the inner and outer edges of the HL Tau disk. The use of the broken power law for the disk’s surface density as well as introducing an increase in the surface density at the boundary between the two segments result in a lower $\chi^2$ value which is shown in Figure \ref{Fig:2alpha}.

\begin{figure*}
    \centering
    \includegraphics[width=18cm,height=5.5cm]{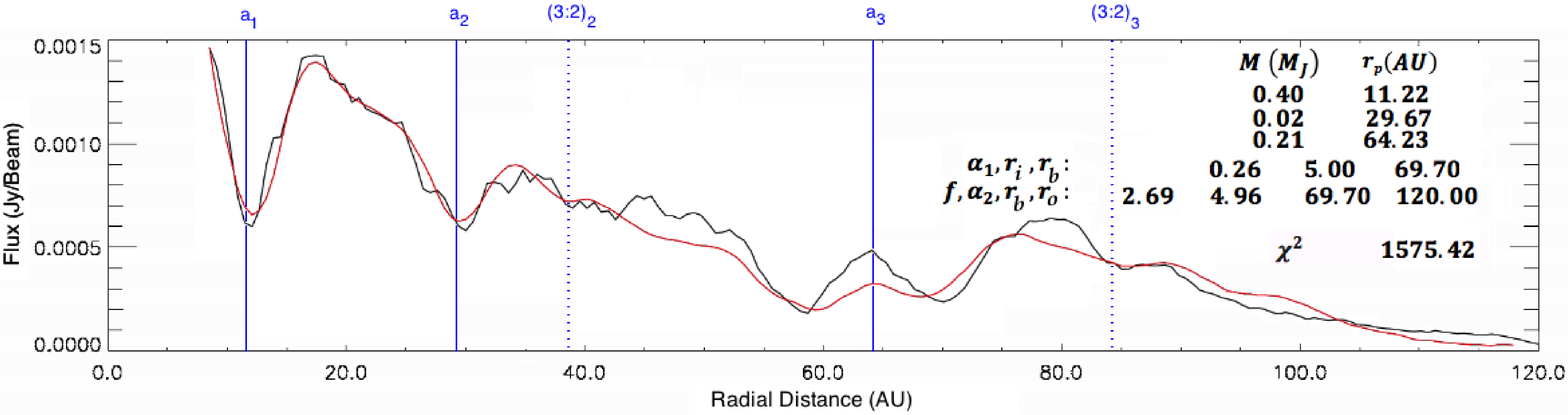}
    \caption{Same as Figure \ref{Fig:2alpha} except that $M_\star$ is changed from 1.3 to 0.55 $M_\odot$.}
    \label{Fig:2alpha_0.55}
\end{figure*}

Figures \ref{Fig:1alpha} and \ref{Fig:2alpha} show our lowest $\chi^2$ results for simulations of disk and planets around a 1.3 $M_\odot$ star with a single and a broken power-law respectively. The lowest $\chi^2$ simulation for the case of $M_\star=0.55~M_\odot$ (broken power-law only) is shown in Figure \ref{Fig:2alpha_0.55}.

\begin{figure*}
    \centering
    \includegraphics[width=18cm,height=8cm]{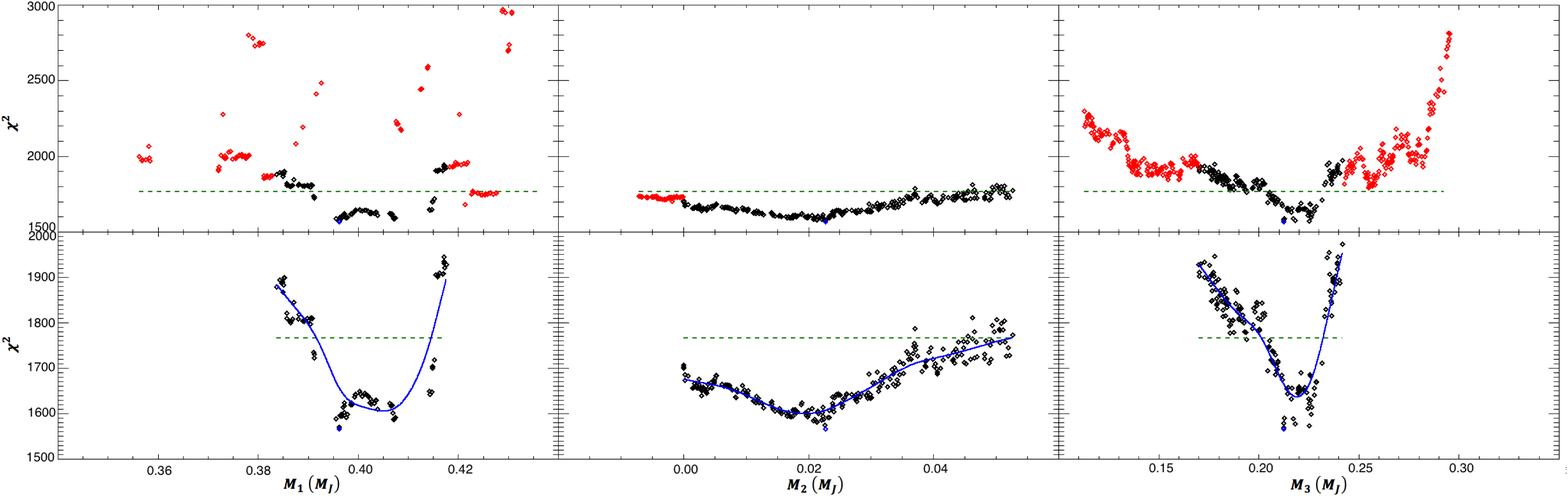}
    \caption[Uncertainty calculations at $2\sigma$ confidence level for the masses of the three planets in the HL Tau disk.]{Uncertainty calculations at $2\sigma$ confidence level for the masses of the three planets in the HL Tau disk shown by Figure \ref{Fig:2alpha_0.55}: $M_1$ (left), $M_2$ (middle), $M_3$ (right). In fitting our spline curve, we exclude the points that fall outside the bowl-shaped part of the $\chi^2$ surface around the minimum (the blue diamond) as well as those that are outside the $2\sigma$ level by more than $10\%$. The excluded points are shown by the red symbols in the top panels. We then fit a spline curve of the lowest possible degree (the blue curve) and note the points where it crosses the $2\sigma$ cut-off (the dashed green line). The difference between the minimum $\chi^2$ and either of those points is taken as the uncertainties for the parameter value.}
    \label{Fig:Uncer_Figs}
\end{figure*}

The $2\sigma$ uncertainties for each parameter are found using the procedure outlined in Section \ref{Sec:uncertainties4}. Figure \ref{Fig:Uncer_Figs} shows uncertainty calculations for the three planet masses. In each case, we fit a polynomial spline curve of the lowest possible degree to the bowl-shaped part of the $\chi^2$ surface and mark the two points where it crosses the $2\sigma$ cut-off. The difference between either of those points and the lowest $\chi^2$ value determines the positive and negative uncertainties.

The best-fit parameters obtained and their uncertainties are shown in Table \ref{Tab:HLTau_params} for the masses and orbital radii of the three planets that we placed in the major gaps of the HL Tau disk for the two different stellar masses used in our simulations. Though some authors have placed two planets in the last two major gaps of the HL Tau disk (at $\sim$ 59 and 70 AU), we are able to reproduce both gaps with a single planet at $\sim$ 64 AU. In agreement with \cite{Dong17} and \cite{Bae17}, we attribute the increase in dust emission at the location of the outermost planet to particles that are trapped in 1:1 MMR with the planet. In the next section, we discuss how the parameters we obtained using our particle-only model compare with those of others that are summarized in Table \ref{Tab:HLTau_params} and whether our method can be used as a first step in modeling complex gas disks such as HL Tau's.


\subsection{Comparison to Other Studies}
\label{Sec:Compare}

A comparison between planetary parameters (masses and orbital radii) that we obtained for the $M_\star=0.55~M_\odot$ case using our simple model of the HL Tau disk shows that, except for the mass of the middle planet which is underestimated by our model, these parameters are comparable to what some authors found using models of higher complexity. In particular, our mass measurements for planets 1 and 3 are similar to those found by \cite{Jin16} who used hydrodynamic gas+dust simulations coupled with 3D radiative transfer calculations. Planet 3's mass is also comparable to what is suggested by \cite{Kanagawa16} through their hydrodynamic simulations. However, those authors note that if the gap they measured was narrower by $\sim$ 2 AU, the mass of the innermost planet would be the same as what \cite{Jin16} found (which is similar to our result).

Planetary masses derived using Equation \ref{Eq:Width_Mass} depend on $r_p$ since gap widths are scaled by the location of their centers. \cite{Kanagawa16} identified the three prominent gaps in the HL Tau disk from the radial profile of the optical depth in band 6 which is offset from the radial profile of the dust brightness temperatures in continuum emission at the locations of the first (D1) and third (D5+D6) planets (see their Figures 1 (a) and (b)). Compared to their plot of the temperature profile of dust, D1 is shifted inward while D5+D6 is shifted outward in optical depth. This means that if $r_p$ was determined from dust temperature, the mass of the planet in D1 would be less than what they report while that of the planet in D5+D6 would be larger. Also they found $r_p$ by taking $(r_{in}+r_{out})/2$ where $r_{in}$ and $r_{out}$ are the inner and outer edges of each gap. This assumes that the gaps are symmetric, but in fact they are slightly asymmetric. This could affect $r_p$ and therefore their calculation of planetary mass from the gap width. Furthermore, \cite{Kanagawa16} also noted that their mass estimates depend strongly on the disk scale height (and hence temperature) as well as dust opacity spectral index, both of which need to be well constrained for the planet mass to be determined accurately. Determining the viscosity parameter is also important when using the formula given by \cite{Kanagawa16} to measure planetary masses, although the dependence is not as strong since $M_p \propto \alpha_{ss}^{1/2}$.

Compared to our results for the planet masses, the masses derived by \cite{Akiyama16} are overestimated. This could be due to the fact that they used gap depths (i.e., Equation \ref{Eq:Depth_Mass}) to find planet masses, and as discussed in Section \ref{Sec:HLTau_Studies}, high signal-to-noise ratio data are required to measure the emission at the bottom of the gap and determine $\Sigma_p/\Sigma_0$ \citep[see][]{Kanagawa16}. Note that both Equations \ref{Eq:Depth_Mass} and \ref{Eq:Width_Mass} apply to gap depth and width in gas emission but assume that they are similar for dust gaps, which is true if gas and dust are well mixed and dust filtration is not strong. However, studies show that even if dust filtration is weak (which is the case for relatively massive disks such as the HL Tau: $M_{disk}=0.07-0.17~M_\odot$; see \cite{Dong15, Jin16}), gas gaps are shallower than dust gaps though the widths remain comparable in gas versus dust. This is because filtration affects gap depths more than their widths \citep[see][]{Dong15, Yen16}. Thus planetary masses are more accurately measured using gap widths (i.e., Equation \ref{Eq:Width_Mass} than gap depths (i.e., Equation \ref{Eq:Depth_Mass}). According to Equation \ref{Eq:Depth_Mass}, shallower gaps result in overestimating planetary mass, which is likely why planetary masses found by \cite{Akiyama16} are larger than those found by ourselves and \cite{Jin16}.

\cite{Tamayo15} did not constrain the masses of the two planets in D1 and D2 since they are dynamically decoupled from the other planets they placed in their simulations. They did, however, determine the limit for the mass of the planet in D5+D6 by letting it grow together with a fourth planet which they placed in D7 ($\sim 90~AU$) under two scenarios: $M_p \lesssim 0.30 ~M_J$ if the two planets are not in MMR and $M_p \gtrsim 0.72 ~M_J$ if they are. We did not put any planet in D7 and leave the investigation of the possibility of additional planets in the HL Tau disk to a future paper, so we shall not comment on how our results compare to theirs other than naively mentioning that in both their four- and five-planet simulations, the masses derived are sub-Jovian (except perhaps where they suggest a lower limit of 0.72 $M_J$ for the outer two planets in MMR in the four-planet case), which is consistent with the other studies mentioned here, ours included. However, we acknowledge that placing additional planets in the system may in fact affect our results, which we defer to future work.

The mass of the central star in the HL Tau system is not well known. Estimates based on the Keplerian velocity of gas \citep[e.g.,][]{Sargent91, Pinte16} or protostellar evolutionary tracks \citep[e.g.,][]{Beckwith90, Gudel08} suggest a star of mass $0.55-1.7~M_\odot$. Therefore, we tried our simulations with a higher stellar mass to see how well our results match those of \cite{Dipierro15} for a stellar mass of $1.3~M_\odot$. Again, the mass of our second planet is significantly lower than theirs while they found the mass of the first planet to be much less than what we did, though the mass of our third planet is comparable to theirs. Since planet masses scale with the stellar mass, which is also true in our simulations when our results for the two different stellar masses are compared against each other, it is not clear why \cite{Dipierro15} found the mass of the first planet to be half that of, for instance, \cite{Jin16} even though the stellar mass is more than doubled.

Our best-fit parameters for the power-law indices of the disk's surface density profile are $\alpha_1=0.26^{+0.02}_{-0.02}$ and $\alpha_2=4.96^{+0.07}_{-0.05}$ where the break occurs at $r_b=69.70^{+0.33}_{-0.12}$. We note that these two values are very different from each other and from what \cite{Yen16} report (see Section \ref{Sec:Fitting4}), partly owing to the fact that we introduced a sudden increase in the disk’s surface density by almost a factor of 3 (i.e., $f=2.69$) where we broke the intensity profile of the disk into the two segments. Furthermore, the \cite{Yen16} values are derived from the the column density of $HCO^+$ gas (though they assumed that dust is tightly coupled to the gas) whereas our model only includes solid dust particles. Nevertheless, our surface density slope in the outer disk is close to the value obtained by \cite{Pinte16} derived from the missing dust mass in each gap. According to their model, the surface density profile of the HL Tau disk has a slope of -3.5 out to about 75 AU but falls off faster in the outer part of the disk. They find the power-law slope in the outer disk to be -4.5, which is similar to what we obtained from our model. They attributed the change in the surface density of dust to two possible reasons: lack of efficient grain growth in the outer disk or the removal of a significant fraction of mm-sized grains from the outer disk via radial migration of dust.

Using a larger stellar mass in our simulations (i.e., $1.3$ instead of $0.55~M_\odot$), we find a similar value for $\alpha_2$ ($=4.68^{+0.06}_{-0.05}$) but $\alpha_1$ is reduced by a factor of 2 ($\alpha_1=0.18^{+0.01}_{-0.01}$). In this case, when using a single power law for the disk’s surface density, we find $\alpha$ to be $\sim 0.20$. We therefore conclude that we would need a more complicated model for the disk surface density to better estimate the power-law indices in the two segments.

Nevertheless, our results explain the three prominent gaps at D1 and D2, and the double gap at D5+D6 and the planetary masses found are similar to the results of others, especially when the stellar mass is $0.55~M_\odot$ while also reproducing some finer gaps, particularly at D3 and D7. In the next section we explain the possible nature of the two narrower gaps seen both in the ALMA image of the HL Tau disk and in our simulations. Therefore, our model is successful in reproducing the observed intensity profile of the HL Tau disk without the need to include certain elements that are necessary to fully study a gas disk.


\subsection{MMR Gaps in the HL Tau Disk}
\label{Sec:HLTau_MMR}

The orbital radii of the planets found by our fitting procedure represent the locations of the three major gaps in the HL Tau image to within uncertainties (where the two gaps made by the outermost planet are considered to be a double gap separated by particles in 1:1 MMR with a planet at $\sim$ 64 AU). A closer look at the observed intensity profile of the HL Tau disk reveals a few other narrower gaps which have motivated some authors to include more planets in their modeling of the HL Tau disk. However, our earlier studies, \cite{Tabeshian16, Tabeshian17},  have shown that not all disk gaps need to contain planetary bodies and that some gaps can, in fact, be made via MMR with a planet that is located outside the gaps and can be used to learn about the hidden planets. We note that our model of the HL Tau disk, which only includes three planets, is able to reproduce some of those narrower gaps as well. In fact, given the locations of the second and the third planets, we argue that the gaps seen at $\sim$38 and $\sim$84 AU roughly corresponding to the locations of the D3 and D7 dark gaps in \cite{Brogan15}, are made by exterior 3:2 MMR with those two planets, respectively. These are shown by the vertical dotted lines in Figures \ref{Fig:2alpha} and \ref{Fig:2alpha_0.55}. Furthermore, the locations of the first and second planets place
them in a 4:1 MMR with each other.

HL Tau is considered a relatively massive disk in which rapid pericenter precession rates alter the location of resonances by an amount roughly given by the ratio of the disk mass to star mass \citep{Tamayo15}. Taking $M_{disk}$ to be $=0.13~M_\odot$ \citep{Kwon11}, this means that the disk is $25\%$ the mass of the star which, according to Equation (10) of \cite{Tamayo15} causes the location of the 3:2 resonance to move by $< 10\%$. However, they point out that, due to the uncertainty in calculating the precession rate, the exact locations of resonances in massive disks are uncertain to within $\sim M_{disk}/M_\star$ as well.

\begin{figure*}
    \centering
    \includegraphics[totalheight=0.34\textheight]{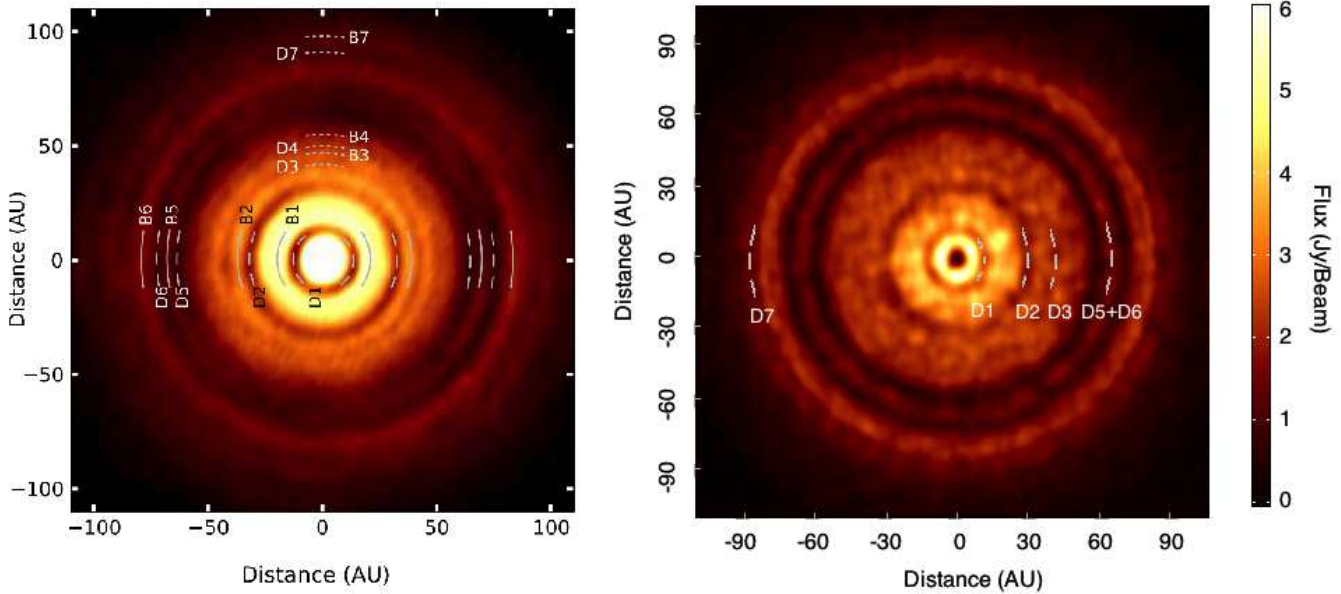}
    \caption[Comparison between ALMA's (deprojected) image of the HL Tau disk with a CASA simulated image drawn from our best-fit parameters.]{Comparison between ALMA’s (deprojected) image of the HL Tau disk \citep{Brogan15} on the left with a CASA simulated image drawn from our best-fit parameters on the right, using a stellar mass of 1.3 $M_\odot$. The dark and bright rings are labeled D1 through D7 and B1 through B7 by \cite{Brogan15}. We use the same notation to mark the locations of the gaps that we believe are sculpted by planets in the HL Tau disk (D1, D2, and D5+D6) as well as the two narrower gaps (D3 and D7) that we believe to be due to MMRs with the embedded planets. The MMR gaps are also marked on Figure \ref{Fig:2alpha} with dotted blue lines. Note that, to make this CASA simulated image, we increased the number of disk particles in our simulation by 10 times for clarity.}
    \label{Fig:ALMA_vs_Sim}
\end{figure*}

In order to visually compare the result of our simulation with ALMA’s image of the HL Tau disk, we make a simulated image using the Common Astronomy Software Applications (CASA) for simulating ALMA observations \citep{CASASim} based on the disk produced with our best-fit parameters. To make the CASA simulated image, we assume that our disk is placed at the HL Tau distance of 130 pc and therefore has the same radial size on the sky. We also assume that the particles are perfect blackbodies at local thermal equilibrium and take the disk’s total flux to be 700 mJy at 1.3 mm \citep{Kwon11}. Stellar radius and effective temperature are $6.0 ~R_\odot$ and $4000 ~K$, respectively \citep{Ruge16}.  We set the image resolution at 35 mas or $\sim$5 AU to match that of ALMA’s observation of the HL Tau disk and use all the 50 available antennas in the 12 m array. We assume that the disk is observed for a total of 4 hr and set the integration time to 10 s per pointing. The R.A. and decl. of the center of the image are $\alpha=04^h31^m38^s.45$ and $\delta=18{\degr}13^\prime59''.0$, J2000 \citep{Tamayo15}. Beam deconvolution is done using CASA's CLEAN algorithm. The result is shown in Figure \ref{Fig:ALMA_vs_Sim}. We adopt the same nomenclature used by \cite{Brogan15} for the dark gaps that we see in our simulations, except that we take the two gaps around the outermost planet to be the same with the planet in the middle.

It must be noted that we are not claiming that the properties of gas-rich disks can be fully determined from simple models that do not incorporate gas and radiation forces. However, based on the ability of our simulations to reproduce the intensity profile of the HL Tau disk, we argue that reasonable matches with observations can be achieved with relatively simple particle-only models of this intrinsically much more complicated gas disk. Therefore, at least as far as understanding the dynamics of the system is involved, we make the case that simple models could be used to extract useful information about the number and properties of possible planets embedded in gas-rich disks which could be used in future, more thorough analyses of these disks.

\section{Summary and Conclusions}
\label{Sec:SummConc4}

Advancements in observing capabilities in recent years have revolutionized our understanding of planet formation and evolution. Interferometric data made available in the mm and sub-mm regime, particularly by ALMA, have provided remarkably detailed images of circumstellar disks with unprecedented angular resolution of a few milliarcseconds. In protoplanetary disks, the structures observed are mostly believed to be due to tidal interactions with unseen planets that clear gaps as they accumulate and then sweep their orbits clear of gas and dust. Therefore, studying such structures would provide insight into the processes involved in the formation and evolution of planets and planetary systems and would help determine some planetary parameters without the need to resolve the planets themselves.

We provided a dynamical model of the HL Tau disk, the most detailed protoplanetary disk structure observed by ALMA to date, without much of the complex physics typically required in modeling gas-rich disks. In particular, we hypothesized that the gas does not dominate the dynamics, and set out to explore whether the radial profile of the HL Tau disk could be recovered using a particle-only model. We were, indeed, able to reproduce the disk’s intensity profile and determine the masses of the planets that could likely be sculpting the most prominent gaps in the HL Tau disk. With the exception of the middle planet’s mass, which is underestimated by our model compared to others, the values we obtained for the masses and radial distances of the three potentially hidden planets in the HL Tau disk orbiting a central star of mass $0.55~M_\odot$ are within the range of parameters quoted in the literature. The planet masses derived from the studies mentioned here are either from more complicated hydrodynamic and SPH simulations or studies that require accurate measurements of disk properties such as scale height (i.e., temperature), viscosity, dust opacity index, and gas-to-dust ratio that are otherwise poorly constrained. Our model is independent of those parameters, which makes it advantageous in arriving rapidly at first estimates for planetary masses without the need to determine the above parameters accurately. However, our results should be taken as first approximations for the masses of the planets; full hydrodynamic models are necessary to study gas-rich disks, such as HL Tau, in more detail. Furthermore, we recovered the tapered-edge of the disk, in which the surface density of the disk changes exponentially beyond the orbit of the outermost planet, as also noted by others, and determined the surface density slope of the disk in the two regions using data from ALMA’s observation at band 7. Another achievement of our model was reproducing a few narrow gaps away from the orbit of the three planets. Whereas the number of planets in the HL Tau disk has remained a matter of debate, our results indicate that at least five gaps can form in the HL Tau disk by including only three planets: the additional gaps are attributed to MMRs with the embedded planets.

Our intention here is not to undermine the importance of hydrodynamic and SPH analyses of gas-rich disks. Though computationally more intensive, such studies are undoubtedly essential in gaining a better understanding of the underlying physics at work in gas disks as sites of planet formation and evolution. However, simpler particle-only models can be used to glean some important information with regard to the dynamics of planet–disk interactions. Such models provide initial conditions to hydrodynamics codes as a first step toward in-depth studies of disk structures, particularly those that are believed to have been formed by unseen planets.

\acknowledgments

We wish to thank the anonymous referee for valuable comments that helped improve this manuscript. This work was supported in part by the Natural Sciences and Engineering Research Council of Canada (NSERC).


\bibliography{main} 

\begin{thebibliography}{}
\expandafter\ifx\csname natexlab\endcsname\relax\def\natexlab#1{#1}\fi

\bibitem[{{Akiyama} {et~al.}(2016){Akiyama}, {Hasegawa}, {Hayashi}, \&
  {Iguchi}}]{Akiyama16}
{Akiyama}, E., {Hasegawa}, Y., {Hayashi}, M., \& {Iguchi}, S. 2016, \apj, 818,
  158

\bibitem[{{ALMA Partnership} {et~al.}(2015){ALMA Partnership}, {Brogan},
  {P{\'e}rez}, {Hunter}, {Dent}, {Hales}, {Hills}, {Corder}, {Fomalont},
  {Vlahakis}, {Asaki}, {Barkats}, {Hirota}, {Hodge}, {Impellizzeri}, {Kneissl},
  {Liuzzo}, {Lucas}, {Marcelino}, {Matsushita}, {Nakanishi}, {Phillips},
  {Richards}, {Toledo}, {Aladro}, {Broguiere}, {Cortes}, {Cortes}, {Espada},
  {Galarza}, {Garcia-Appadoo}, {Guzman-Ramirez}, {Humphreys}, {Jung}, {Kameno},
  {Laing}, {Leon}, {Marconi}, {Mignano}, {Nikolic}, {Nyman}, {Radiszcz},
  {Remijan}, {Rod{\'o}n}, {Sawada}, {Takahashi}, {Tilanus}, {Vila Vilaro},
  {Watson}, {Wiklind}, {Akiyama}, {Chapillon}, {de Gregorio-Monsalvo}, {Di
  Francesco}, {Gueth}, {Kawamura}, {Lee}, {Nguyen Luong}, {Mangum}, {Pietu},
  {Sanhueza}, {Saigo}, {Takakuwa}, {Ubach}, {van Kempen}, {Wootten},
  {Castro-Carrizo}, {Francke}, {Gallardo}, {Garcia}, {Gonzalez}, {Hill},
  {Kaminski}, {Kurono}, {Liu}, {Lopez}, {Morales}, {Plarre}, {Schieven},
  {Testi}, {Videla}, {Villard}, {Andreani}, {Hibbard}, \&
  {Tatematsu}}]{Brogan15}
{ALMA Partnership}, {Brogan}, C.~L., {P{\'e}rez}, L.~M., {et~al.} 2015, \apjl,
  808, L3

\bibitem[{{Andrews} \& {Williams}(2007)}]{Andrews07b}
{Andrews}, S.~M., \& {Williams}, J.~P. 2007, \apj, 659, 705

\bibitem[{Bae {et~al.}(2017)Bae, Zhu, \& Hartmann}]{Bae17}
Bae, J., Zhu, Z., \& Hartmann, L. 2017, \apj, 850, 201

\bibitem[{{Beckwith} {et~al.}(1990){Beckwith}, {Sargent}, {Chini}, \&
  {Guesten}}]{Beckwith90}
{Beckwith}, S.~V.~W., {Sargent}, A.~I., {Chini}, R.~S., \& {Guesten}, R. 1990,
  \aj, 99, 924

\bibitem[{{Birnstiel} \& {Andrews}(2014)}]{Birnstiel14}
{Birnstiel}, T., \& {Andrews}, S.~M. 2014, \apj, 780, 153

\bibitem[{{Collins} {et~al.}(2009){Collins}, {Grady}, {Hamaguchi},
  {Wisniewski}, {Brittain}, {Sitko}, {Carpenter}, {Williams}, {Mathews},
  {Williger}, {van Boekel}, {Carmona}, {Henning}, {van den Ancker}, {Meeus},
  {Chen}, {Petre}, \& {Woodgate}}]{Collins09}
{Collins}, K.~A., {Grady}, C.~A., {Hamaguchi}, K., {et~al.} 2009, \apj, 697,
  557

\bibitem[{{Dipierro} {et~al.}(2015){Dipierro}, {Price}, {Laibe}, {Hirsh},
  {Cerioli}, \& {Lodato}}]{Dipierro15}
{Dipierro}, G., {Price}, D., {Laibe}, G., {et~al.} 2015, \mnras, 453, L73

\bibitem[{{Dong} {et~al.}(2017){Dong}, {Li}, {Chiang}, \& {Li}}]{Dong17}
{Dong}, R., {Li}, S., {Chiang}, E., \& {Li}, H. 2017, \apj, 843, 127

\bibitem[{{Dong} {et~al.}(2015){Dong}, {Zhu}, \& {Whitney}}]{Dong15}
{Dong}, R., {Zhu}, Z., \& {Whitney}, B. 2015, \apj, 809, 93

\bibitem[{{Duffell} \& {MacFadyen}(2013)}]{Duffell13}
{Duffell}, P.~C., \& {MacFadyen}, A.~I. 2013, \apj, 769, 41

\bibitem[{{Fouchet} {et~al.}(2010){Fouchet}, {Gonzalez}, \&
  {Maddison}}]{Fouchet10}
{Fouchet}, L., {Gonzalez}, J.-F., \& {Maddison}, S.~T. 2010, \aap, 518, A16

\bibitem[{{Fouchet} {et~al.}(2007){Fouchet}, {Maddison}, {Gonzalez}, \&
  {Murray}}]{Fouchet07}
{Fouchet}, L., {Maddison}, S.~T., {Gonzalez}, J.-F., \& {Murray}, J.~R. 2007,
  \aap, 474, 1037

\bibitem[{{Fung} {et~al.}(2014){Fung}, {Shi}, \& {Chiang}}]{Fung14}
{Fung}, J., {Shi}, J.-M., \& {Chiang}, E. 2014, \apj, 782, 88

\bibitem[{{Gladman}(1993)}]{Gladman93}
{Gladman}, B. 1993, \icarus, 106, 247

\bibitem[{{G{\"u}del} {et~al.}(2008){G{\"u}del}, {Arzner}, {Audard}, {Bouvier},
  {Briggs}, {Dougados}, {Feigelson}, {Franciosini}, {Glauser}, {Grosso},
  {Guieu}, {M{\'e}nard}, {Micela}, {Monin}, {Montmerle}, {Padgett}, {Palla},
  {Pillitteri}, {Preibisch}, {Rebull}, {Scelsi}, {Silva}, {Skinner}, {Stelzer},
  \& {Telleschci}}]{Gudel08}
{G{\"u}del}, M., {Arzner}, K., {Audard}, M., {et~al.} 2008, in Astronomical
  Society of the Pacific Conference Series, Vol. 384, 14th Cambridge Workshop
  on Cool Stars, Stellar Systems, and the Sun, ed. G.~{van Belle}, 65

\bibitem[{{Haghighipour} \& {Boss}(2003)}]{Haghighipour03}
{Haghighipour}, N., \& {Boss}, A.~P. 2003, \apj, 583, 996

\bibitem[{{Hartmann} {et~al.}(1998){Hartmann}, {Calvet}, {Gullbring}, \&
  {D'Alessio}}]{Hartmann98}
{Hartmann}, L., {Calvet}, N., {Gullbring}, E., \& {D'Alessio}, P. 1998, \apj,
  495, 385

\bibitem[{{Jin} {et~al.}(2016){Jin}, {Li}, {Isella}, {Li}, \& {Ji}}]{Jin16}
{Jin}, S., {Li}, S., {Isella}, A., {Li}, H., \& {Ji}, J. 2016, \apj, 818, 76

\bibitem[{{Kanagawa} {et~al.}(2015){Kanagawa}, {Muto}, {Tanaka}, {Tanigawa},
  {Takeuchi}, {Tsukagoshi}, \& {Momose}}]{Kanagawa15}
{Kanagawa}, K.~D., {Muto}, T., {Tanaka}, H., {et~al.} 2015, \apjl, 806, L15

\bibitem[{{Kanagawa} {et~al.}(2016){Kanagawa}, {Muto}, {Tanaka}, {Tanigawa},
  {Takeuchi}, {Tsukagoshi}, \& {Momose}}]{Kanagawa16}
---. 2016, \pasj, 68, 43

\bibitem[{{Kwon} {et~al.}(2011){Kwon}, {Looney}, \& {Mundy}}]{Kwon11}
{Kwon}, W., {Looney}, L.~W., \& {Mundy}, L.~G. 2011, \apj, 741, 3

\bibitem[{{Levenberg}(1944)}]{Levenberg44}
{Levenberg}, K. 1944, {\textit{A Method for the Solution of Certain Non-Linear
  Problems in Least Squares}, The Quarterly of Applied Mathematics, 2, 164}

\bibitem[{{Maddison} {et~al.}(2007){Maddison}, {Fouchet}, \&
  {Gonzalez}}]{Maddison07}
{Maddison}, S.~T., {Fouchet}, L., \& {Gonzalez}, J.-F. 2007, \apss, 311, 3

\bibitem[{{Marquardt}(1944)}]{Marquardt44}
{Marquardt}, D. 1944, {\textit{An algorithm for least-squares estimation of
  nonlinear parameters}, Journal of the Society for Industrial and Applied
  Mathematics, 11, 431}

\bibitem[{{McCaughrean} \& {O'dell}(1996)}]{McCaughrean96}
{McCaughrean}, M.~J., \& {O'dell}, C.~R. 1996, \aj, 111, 1977

\bibitem[{{McMullin} {et~al.}(2007){McMullin}, {Waters}, {Schiebel}, {Young},
  \& {Golap}}]{CASASim}
{McMullin}, J.~P., {Waters}, B., {Schiebel}, D., {Young}, W., \& {Golap}, K.
  2007, in Astronomical Society of the Pacific Conference Series, Vol. 376,
  Astronomical Data Analysis Software and Systems XVI, ed. R.~A. {Shaw},
  F.~{Hill}, \& D.~J. {Bell}, 127

\bibitem[{{Murray} \& {Dermott}(1999)}]{Murray99}
{Murray}, C.~D., \& {Dermott}, S.~F. 1999, {Solar System Dynamics (Cambridge:
  Cambridge Univ. Press)}

\bibitem[{{Nelder} \& {Mead}(1964)}]{Nelder64}
{Nelder}, J.~A., \& {Mead}, R. 1964, The Computer Journal, 7, 308

\bibitem[{NRAO(2014)}]{NRAO14}
NRAO. 2014, Birth of Planets Revealed in Astonishing Detail in ALMA's "Best
  Image Ever", National Radio Astronomy Observatory (Press Release),
  https://public.nrao.edu/news/2013-10-01-13-11-51-2/

\bibitem[{{Paardekooper} \& {Mellema}(2004)}]{Paardekooper04}
{Paardekooper}, S.-J., \& {Mellema}, G. 2004, \aap, 425, L9

\bibitem[{{Pinte} {et~al.}(2016){Pinte}, {Dent}, {M{\'e}nard}, {Hales}, {Hill},
  {Cortes}, \& {de Gregorio-Monsalvo}}]{Pinte16}
{Pinte}, C., {Dent}, W.~R.~F., {M{\'e}nard}, F., {et~al.} 2016, \apj, 816, 25

\bibitem[{{Press} {et~al.}(1992){Press}, {Teukolsky}, {Vetterling}, \&
  {Flannery}}]{Press92}
{Press}, W.~H., {Teukolsky}, S.~A., {Vetterling}, W.~T., \& {Flannery}, B.~P.
  1992, {Numerical Recipes in C. The Art of Scientific Computing (Cambridge:
  Cambridge Univ. Press)}

\bibitem[{{Price} {et~al.}(2017){Price}, {Wurster}, {Nixon}, {Tricco},
  {Toupin}, {Pettitt}, {Chan}, {Laibe}, {Glover}, {Dobbs}, {Nealon}, {Liptai},
  {Worpel}, {Bonnerot}, {Dipierro}, {Ragusa}, {Federrath}, {Iaconi},
  {Reichardt}, {Forgan}, {Hutchison}, {Constantino}, {Ayliffe}, {Mentiplay},
  {Hirsh}, \& {Lodato}}]{Price17}
{Price}, D.~J., {Wurster}, J., {Nixon}, C., {et~al.} 2017, ArXiv e-prints,
  arXiv:1702.03930

\bibitem[{{Rein} \& {Liu}(2012)}]{Rein12}
{Rein}, H., \& {Liu}, S.-F. 2012, \aap, 537, A128

\bibitem[{{Ruge} {et~al.}(2016){Ruge}, {Flock}, {Wolf}, {Dzyurkevich},
  {Fromang}, {Henning}, {Klahr}, \& {Meheut}}]{Ruge16}
{Ruge}, J.~P., {Flock}, M., {Wolf}, S., {et~al.} 2016, \aap, 590, A17

\bibitem[{{Sargent} \& {Beckwith}(1991)}]{Sargent91}
{Sargent}, A.~I., \& {Beckwith}, S.~V.~W. 1991, \apjl, 382, L31

\bibitem[{{Shakura} \& {Sunyaev}(1973)}]{Shakura73}
{Shakura}, N.~I., \& {Sunyaev}, R.~A. 1973, \aap, 24, 337

\bibitem[{{Tabeshian} \& {Wiegert}(2016)}]{Tabeshian16}
{Tabeshian}, M., \& {Wiegert}, P.~A. 2016, \apj, 818, 159

\bibitem[{{Tabeshian} \& {Wiegert}(2017)}]{Tabeshian17}
---. 2017, \apj, 847, 24

\bibitem[{{Takahashi} \& {Inutsuka}(2014)}]{Takahashi14}
{Takahashi}, S.~Z., \& {Inutsuka}, S.-i. 2014, \apj, 794, 55

\bibitem[{{Tamayo} {et~al.}(2015){Tamayo}, {Triaud}, {Menou}, \&
  {Rein}}]{Tamayo15}
{Tamayo}, D., {Triaud}, A.~H.~M.~J., {Menou}, K., \& {Rein}, H. 2015, \apj,
  805, 100

\bibitem[{{Testi} {et~al.}(2015){Testi}, {Skemer}, {Henning}, {Bailey},
  {Defr{\`e}re}, {Hinz}, {Leisenring}, {Vaz}, {Esposito}, {Fontana}, {Marconi},
  {Skrutskie}, \& {Veillet}}]{Testi15}
{Testi}, L., {Skemer}, A., {Henning}, T., {et~al.} 2015, \apjl, 812, L38

\bibitem[{{van der Marel} {et~al.}(2013){van der Marel}, {van Dishoeck},
  {Bruderer}, {Birnstiel}, {Pinilla}, {Dullemond}, {van Kempen}, {Schmalzl},
  {Brown}, {Herczeg}, {Mathews}, \& {Geers}}]{vanderMarel13}
{van der Marel}, N., {van Dishoeck}, E.~F., {Bruderer}, S., {et~al.} 2013,
  Science, 340, 1199

\bibitem[{{Weidenschilling}(1977)}]{Weidenschilling77}
{Weidenschilling}, S.~J. 1977, \mnras, 180, 57

\bibitem[{{Wisdom} \& {Holman}(1991)}]{Wisdom91}
{Wisdom}, J., \& {Holman}, M. 1991, \aj, 102, 1528

\bibitem[{{Yen} {et~al.}(2016){Yen}, {Liu}, {Gu}, {Hirano}, {Lee},
  {Puspitaningrum}, \& {Takakuwa}}]{Yen16}
{Yen}, H.-W., {Liu}, H.~B., {Gu}, P.-G., {et~al.} 2016, \apjl, 820, L25

\bibitem[{{Zhang} {et~al.}(2015){Zhang}, {Blake}, \& {Bergin}}]{Zhang15}
{Zhang}, K., {Blake}, G.~A., \& {Bergin}, E.~A. 2015, \apjl, 806, L7

\end{thebibliography}

\end{document}